\documentclass[twocolumn,superscriptaddress]{revtex4-2}
\usepackage{gnuplottex}
\usepackage{mathtools}
\usepackage{float}
\usepackage{array}
\usepackage[english]{babel}
\usepackage[utf8]{inputenc}
\usepackage{amsmath}
\usepackage[table]{xcolor}
\usepackage{amsfonts}
\usepackage{graphicx}
\usepackage{dsfont}
\usepackage[colorlinks=true, linkcolor=blue, citecolor=dred]{hyperref}
\usepackage{algorithm}
\usepackage[noend]{algpseudocode}
\usepackage{bbold}
\usepackage{mathtools}
\usepackage{amssymb}
\usepackage{bm}
\usepackage{bbm}
\usepackage{tabularx, booktabs}
\usepackage{xspace}
\usepackage{listings}
\usepackage[noend]{algpseudocode}
\newcolumntype{Y}{>{\centering\arraybackslash}X}
\definecolor{dgreen}{rgb}{0,.5,0}
\definecolor{dblue}{rgb}{0,0,.5}
\definecolor{dred}{rgb}{0.5,0,.5}

\newcommand\reallywidehat[1]{%
\savestack{\tmpbox}{\stretchto{%
  \scaleto{%
    \scalerel*[\widthof{\ensuremath{#1}}]{\kern-.6pt\bigwedge\kern-.6pt}%
    {\rule[-\textheight/2]{1ex}{\textheight}}
  }{\textheight}%
}{0.5ex}}%
\stackon[1pt]{#1}{\tmpbox}%
}
\begin{document}

\title{The extended star graph as a light-harvesting-complex prototype: \\ excitonic absorption speedup by peripheral energy defect tuning 
} 
\author{Saad Yalouz} 
\email{yalouzsaad@gmail.com}
\affiliation{Laboratoire de Chimie Quantique, Institut de Chimie,
CNRS/Université de Strasbourg, 4 rue Blaise Pascal, 67000 Strasbourg, France} 
\author{Vincent Pouthier}
\affiliation{Institut UTINAM,  Universit\'{e} de Franche-Comt\'{e}, CNRS UMR 6213, 25030 Besan\c {c}on, France}

\begin{abstract}
We study the quantum dynamics of a photo-excitation uniformly distributed at the periphery of an extended star network (with $N_B$ branches of length $L_B$).  
More specifically, we address here the question of the energy absorption at the core of the network and how this process can be improved (or not) by the inclusion of peripheral defects with a tunable energy amplitude $\Delta$.  
Our numerical simulations reveal the existence of optimal value of energy defect $\Delta^*$ which depends on the network architecture. 
Around this value, the absorption process presents a strong speedup (\textit{i.e.} reduction of the absorption time) provided that  $L_B \leq L_B^*$ with $L_B^* \approx 12.5/\ln(N_B) $.
Analytical/numerical developments are then conducted to interpret this feature. 
We show that the origin of this speedup takes place in the hybridization of two upper-band excitonic eigenstates.
This hybridization is important when $L_B \leq L_B^*$ and vanishes almost totally when  $L_B > L_B^*$. 
These structural rules  we draw here could represent a potential guide for the practical design of molecular nano-network dedicated to the realisation of efficient photo-excitation absorption.
\end{abstract}

\maketitle

\section{Introduction} 

In natural photosynthesis, the solar energy is absorbed by a non-trivial arrangement of chlorophyll molecules called  ``light-harvesting complex'' (LHC)~\cite{barber07,cheng09,scholes11}. After absorption, the energy of light is efficiently transferred to a reaction center where the adenosine triphosphate (ATP) is finally produced. The energy released by the hydrolysis of ATP represents the chemical fuel at the origin of life in the plant kingdom. Because the sun is a clean energy source with rich reserves, it turns out that the solar energy is one of the best strategies to overcome the current energy crisis~\cite{nocera12,nocera06}. As a consequence, photosynthesis has recently attracted a large amount of research interest. More precisely, a special attention has been paid to elaborate artificial LHCs that could be able to mimic natural photosynthesis to efficiently convert the energy of light into chemical fuel~\cite{wang22,kundu17,jiang17,osella21}. 

In particular, it has been suggested that exciton-mediated energy transport could be exploited in dendrimers to design artificial LHCs~\cite{mukamel97}. A dendrimer is an engineered polymer whose hyperbranched structure at nanoscale looks like the fractal patterns that occur in the plant kingdom~\cite{vogtle09,astruc10,frechet94,bosman99}. It consists of several dendritic branches, called dendrons, that emanate out from a central core. Each dendron is formed by long molecular chains organized in a self-similar fashion. It exhibits branching points where the chain splits into two or three chains, increasing the generation number, and its periphery is occupied by functional terminal groups. To obtain a LHC, the main idea consists in the functionalization of the terminal groups by chromophores that favor light harvesting. The absorbed light yields Frenkel excitons that propagate along the dendrons and converge toward the central core where chemical fuel is finally produced~\cite{ harigaya99,delgado02,supritz08,tretiak98,nakano00,nakano01,crabtree07}. Note that dendrimers are not the only systems that have been considered. Many alternatives have been elaborated owing to the prowess of the chemical engineering, such as porphyrin arrays~\cite{wang12,waki12,miyasaka11} , organic nanocrystals~\cite{sun19,sun18} or LHC based on non-covalent interactions~\cite{peng16,shulov15,kumar21}, to cite but a few examples. 

From a theoretical point of view, the exciton propagation in an artificial LHC can be viewed as a physical realization of a continuous time quantum walk~\cite{mulken06a,benedetti21} on a network that exhibits a complex architecture, defects (the light absorbers) and a trap (the reaction center). 
Therefore, in order to judge the efficiency of the light capture process, it is of fundamental importance to understand the interaction between these three ingredients. Specifically, the question arises as to how to design the network so that these key ingredients prevent excitonic localisation and instead promote, as far as possible, the propagation of energy towards the trap for an efficient absorption process. 

Indeed, as shown by Mulken et al.~\cite{mulken06b}, localization processes may first result from the complex nature of the network that favors the occurrence of highly degenerate excitonic eigenstates. Therefore, when the excitonic quantum state initially expands over few degenerate eigenstates, specific quantum self-interferences arise. The propagation of the exciton is thus stopped so that the exciton remains confined in the neighborhood of the excited region on the network. Such a feature has been reported in many networks such as compact dendrimers~\cite{mulken06c}, star graphs~\cite{xu09}, and Apollonian networks~\cite{xu08}. But the architecture of the network may also induce localization even if the exciton occupies initially a non degenerate eigenstate.  This feature has been observed in an extended dendrimer, when the initial wave function is uniformly distributed over the periphery~\cite{pouthier13}. In that case quantum interferences arise because multiple scatterings occur each time the exciton tunnels from one generation to another resulting in a localization process when the generation number exceeds a critical value.

 Then, as in solid-state physics, the localization may arise when the network is perturbed by energetic defects~\cite{maradudin66}. Indeed, a single defect breaks the symmetry of the network resulting in the occurrence of an excitonic wave function that exponentially localizes in the neighborhood of the defect. Nevertheless, more subtle situations may arise when several defects are present. Indeed, in accordance with the well-known concept of localization transition due to Anderson~\cite{anderson58,kramer93}, the defects may act as a negative ingredient which prevents the excitonic delocalization as observed in linear chains~\cite{yin08}, discrete rings~\cite{mulken07}, and binary trees~\cite{rebentrost09}. By contrast, in other cases, the presence of defects may behave as a positive ingredient that enhances the delocalized nature of the exciton. This feature has been observed in tree graphs similar to dendrimers where a weak disorder yields extended states through fluctuation-enabled resonances between states that initially may appear to be localized~\cite{aizenman11,aizenman13}.

Finally, the interaction with traps drastically affects the efficiency of the excitonic propagation~\cite{celardo12,celardo16,celardo17,yalouz18,yalouz20}. Originating in the coupling with an external continuum, the trapping effect is usually addressed using a non-Hermitian exciton Hamiltonian.  The real parts of the complex eigenvalues of the Hamiltonian define the excitonic energies, whereas the imaginary parts specify the energy widths, i.e., the decay rates. In that case, it has been shown that a general phenomenon called superradiance transition may arise~\cite{celardo09}. Indeed, when the exciton-trap coupling is weak, all the excitonic eigenstates exhibit quite similar decay rates. However, as the coupling increases, an eigenstate restructuring takes place. Only a few short-lived states, called superradiant states, exhibit cooperatively enhanced decay rates. These states are accompanied by subradiant eigenstates which represent long-lived states almost decoupled from the traps. When the exciton-trap coupling becomes sufficiently strong, the superradiant states localize on the traps so that the excitonic transfer to those specific sites is drastically hindered and the trapping process loses in efficiency.

In this paper, we present a theoretical work to illustrate how the interplay between the complex architecture, the presence of defects and the use of a trap could be exploited in an extended star to design an efficient LHC. The star graph is one of the most regular structures in graph theory. Organized around a central core, it exhibits the local tree structure of irregular and complex networks. However, its topology remains sufficiently simple so that analytical calculations can be carried out. To proceed, one considers that the periphery of the extended star graph is functionalized by energetic defects whereas the core is occupied by a trap. The absorption of light by the defects brings the exciton in an initial state uniformly delocalized over the peripheral sites whereas the trap is responsible for the exciton decay. Therefore, depending on the structure of the graph, it will be shown that if the energetic defects are judiciously chosen, the initial state localized at the periphery may hybridize with a state localized on the core. As a consequence, a speedup of the excitonic propagation is observed and a quite efficient artificial LHC is finally obtained. 

The paper is organized as follows, in Sec.~\ref{sec:theory} the extended star graph is introduced and the exciton Hamiltonian is defined.
Then the relevant observables required for characterizing the dynamics and the absorption process are described.
In Sec.~\ref{sec:numerical}, a numerical analysis is performed for characterizing the absorption process.
Finally, in Sec.~\ref{sec:discussion} the results are discussed and interpreted using a semi-analytical approach.

\section{Theoretical Background}\label{sec:theory}

\subsection{Model Hamiltonian}

\begin{figure*}
\centering
\includegraphics[width=0.85\textwidth]{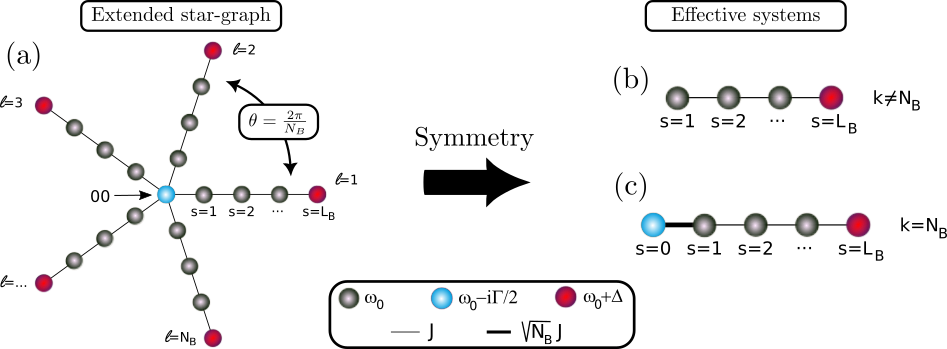}
\caption{\textbf{Illustration of the extended star graph and the problem reduction based on rotational symmetries.} \textbf{(a) }The original network composed of a central site connected to  $N_B$ branches with $L_B$ sites each. The total number of sites is $1+N_BN_L$. We consider here the presence of a trap on the central site (represented in blue), and $N_B$ energetic defects located on the periphery of the star (represented in red). A rotational symmetry of angle $\theta = 2\pi/N_B$ is present. This symmetry can be used to   reduce the complexity of the whole system into a series of effective sub-graphs with \textbf{(b)}~$N_B-1$ effective chains modeling the Hamiltonian given in Eq.~(\ref{eq:k_diff_NB}), and  \textbf{(c)}~ the particular case of the effective chain modeling the Hamiltonian given in Eq.~(\ref{eq:k_eq_NB}). This last system (\textit{i.e.} Hamiltonian) is the one we employ to describe the quantum dynamics of an initial excitation uniformly delocalized on the periphery of the original network.  }
\label{fig:star}
\end{figure*}

As shown in Fig.~\ref{fig:star}, we consider the extended star graph formed by $N_B$ branches that emanate out from a central node. Each branch $\ell =1,2,...,N_B$ corresponds to a finite size chain whose nodes are labeled by the index $s=1,2,...,L_B$. The central node, denoted $(\ell=0,s=0)$, is connected to the side site $s=1$ of each branch. All the nodes are identical excluding those of the periphery of the star, the terminal groups being functionalized by energetic defects, as well as the core site that is occupied by a trap.



In this network, we are interested in the motion of an exciton whose quantum dynamics is described according to a standard tight-binding model. Within this model, each site $(\ell,s)$ is occupied by a molecular subunit whose internal dynamics is described by a two-level system. Let $|\ell,s\rangle$ stand for the state in which the $(\ell,s)$th two-level system occupies its first excited state, the other two-level systems remaining in their ground state. The vacuum state $|\oslash\rangle$ describes all the two-level systems in their ground state. Let $\omega_0$ denote the Bohr frequency of the two-level systems except those of the periphery and of the core of the graph. The terminal groups are occupied by energetic defects whose Bohr frequency is shifted by an among $\Delta$ according to $\omega_0+\Delta$. The central node is occupied by a trap that is responsible for the irreversible decay of the exciton. It is characterized by a complex self-energy $\omega_0-i\Gamma/2$ where $\Gamma$ defines the exciton decay rate~\cite{sokolov92,sokolov89}. Finally, the exciton is able to tunnel from one node to another according to the hopping constant $J$ that connects the linked nodes.

Within these notations, the Hamiltonian that governs the one-exciton dynamics is defined as 
(with the convention $\hbar = 1$)
\begin{eqnarray}
 \hat{H} &=&\left(\omega_0-i\frac{\Gamma}{2} \right) |00\rangle \langle 00|+\sum_{\ell=1}^{N_B}\sum_{s=1}^{L_B} (\omega_0+\Delta \delta_{s L_B}) |\ell s\rangle \langle \ell s |  \nonumber \\
         &+&\sum_{\ell=1}^{N_B} J  ( |00\rangle \langle \ell 1 |+|\ell 1\rangle \langle 00 |) \nonumber \\
         &+&\sum_{\ell=1}^{N_B}\sum_{s=1}^{L_B-1} J  ( |\ell s\rangle \langle \ell s+1 |+|\ell s+1\rangle \langle \ell s |). 
\end{eqnarray}

\subsection{Block diagonal representation}

For describing the exciton eigenstates, we can take advantage of the fact that the Hamiltonian $\hat{H}$ is invariant under the discrete rotation of angle $\theta= 2 \pi /N_B$ and centered on the core of the star (see Fig.~\ref{fig:star}.(a)). Consequently, its diagonalization is greatly simplified when one works with an intermediate basis that involves the state localized on the core $|00\rangle$  and a set of orthogonal Bloch states $|\chi^{(k)}_s \rangle $ with $s=1,2,...,L_B$ and $k = 1,2,...,N_B$. A Bloch state is defined as
\begin{equation}
|\chi^{(k)}_s \rangle=\frac{1}{\sqrt{N_B}} \sum_{\ell=1}^{N_B} e^{-ik\ell \theta} |\ell s \rangle.
\end{equation}
Within this basis, it turns out that $k$ is a good quantum number so that the Hamiltonian $\hat{H}$ becomes block diagonal. It is expressed as a direct sum 
$\hat{H}=\hat{H}^{(1)} \oplus \hat{H}^{(2)}...\oplus \hat{H}^{(N_B)}$  where $\hat{H}^{(k)}$ is the block Hamiltonian associated to the quantum number $k$. Therefore, two situations arise depending on the value of the integer $k$.

For all $k \neq N_B$, all the block $\hat{H}^{(k)}$ are identical. They are expressed as 
\begin{equation}\label{eq:k_diff_NB}
\begin{split}
\hat{H}^{(k \neq N_B)} &= \sum_{s=1}^{L_B} (\omega_0+\Delta \delta_{s L_B}) |\chi^{(k)}_s \rangle \langle \chi^{(k)}_s | \\
                       &+  \sum_{s=1}^{L_B-1} J  ( |\chi^{(k)}_{s+1} \rangle \langle \chi^{(k)}_s |+|\chi^{(k)}_{s} \rangle \langle \chi^{(k)}_{s+1}|). 
\end{split}
\end{equation}
As illustrated in Fig.~\ref{fig:star}(b), they correspond to a tight-binding Hamiltonian on a finite size chain formed by the sites $s=1,2,...,L_B$. They involve the states $|\chi^{(k)}_{s} \rangle $ but they do not involve the state $|00\rangle$ localized on the core of the extended star graph. Therefore, $\hat{H}^{(k \neq N_B)}$ acts in a Hilbert 
space $E^{(k \neq N_B)}$  whose dimension reduces to $L_B$. Each site is characterized by a self-energy $\omega_0$ excluding the side site $s=L_B$ whose self-energy is shifted by an among $\Delta$.


For $k=N_B$, a different situation occurs. Indeed, in that case, the Hamiltonian $\hat{H}^{(N_B)}$ is defined as 
\begin{equation} \label{eq:k_eq_NB}
\begin{split}
    \hat{H}^{(N_B)} &= \left(\omega_0-i\frac{\Gamma}{2} \right) |00\rangle \langle 00| \\
              &+ \sum_{s=1}^{L_B} (\omega_0+\Delta \delta_{s L_B}) |\chi^{(N_B)}_s \rangle \langle \chi^{(N_B)}_s | \\
              &+ \sqrt{N_B} J (|00\rangle \langle \chi_1^{(N_B)} |+|\chi_1^{(N_B)} \rangle \langle 00| ) \\
              &+ \sum_{s=1}^{L_B-1} J ( |\chi^{(N_B)}_{s+1} \rangle \langle \chi^{(N_B)}_s |+|\chi^{(N_B)}_{s} \rangle \langle \chi^{(N_B)}_{s+1}|).
\end{split}
\end{equation} 
As previously, $\hat{H}^{(N_B)}$ defines a tight-binding Hamiltonian on a finite size chain (see Fig.~\ref{fig:star}(c)). This chain involves the sites $s=0,1,...,L_B$ associated to the states $|00\rangle$ (the exciton is located on the core of the extended star), $|\chi_1^{(N_B)} \rangle$ (the exciton is uniformly delocalized over the sites $s=1$ of the branches of the extended star),  $|\chi_2^{(N_B)} \rangle$  (the exciton is uniformly delocalized over the sites $s=2$ of the branches of the extended star), ... and so on. When compared with what happens when $k \neq N_B$, the finite size chain exhibits three defects. First, a “complex” defect is localized on the side site $s=0$ whose self-energy $\omega_0-i\Gamma/2$ is modified by the presence of the trap. Second, the strength of the link between the side sites $s=0$ and $s=1$ is equal to $\sqrt{N_B}J$ whereas in the core of the chain this strength reduces to $J$. Finally, an energetic defect is localized on the side site $s=L_B$ whose self-energy is shifted by an among $\Delta$ when compared with that of the other sites. Note that $ \hat{H}^{(N_B)}$ acts in a Hilbert space whose dimension is equal to $L_B+1$. 


Within this block diagonal representation of the exciton Hamiltonian $\hat{H}$, the corresponding Schrödinger equation can be solved numerically to determine the eigenvalues $\{\bar{\omega}_\mu^{(k)} \}$ and the associated eigenvectors  $\{ |\varphi_\mu^{(k)} ) \}$ labeled by the indexes $k$ and $\mu$. From the knowledge of these eigen-properties, one can compute in principle all the observables needed for characterizing the dynamics. 

However, in the present work, we shall focus our attention to a particular situation in which the interaction with an external field, such as an electromagnetic field, will be assumed to bring the exciton in a state that is uniformly delocalized over the periphery of the star. Consequently, the exciton dynamics is confined in the $k=N_B$ subspace. Restricting our attention to that subspace, the notations are simplified as follows. First, one disregards the index $k$ and introduce the Hamiltonian  $\mathcal{\hat{H}}=\hat{H}^{(N_B)}$. Then, the basis vectors are renamed as
\begin{equation}
  |s) = \left\{
      \begin{aligned}
        |00\rangle   &     \ \ \textrm{if} & s=0 \\
        |\chi_s^{(N_B)}\rangle &    \ \ \textrm{if} & s>0. \\
      \end{aligned}
    \right.
\end{equation}
Within these simplified notations, the restriction of the Hamiltonian in the $k=N_B$ subspace is finally rewritten 
\begin{equation} \label{eq:reduced_hamiltonian}
\begin{split}
 \hat{\mathcal{H}} &= \sum_{s=0}^{L_B} \left(\omega_0-i\frac{\Gamma}{2} \delta_{s0}+\Delta \delta_{s L_B} \right)  | s) (s|  \\
                   &+ \sqrt{N_B} J ( |0)(1| + |1)(0|) \\
                   &+ \sum_{s=1}^{L_B-1} J  (|s)(s+1| + |s+1)(s|).   
\end{split}
\end{equation}

 \subsection{Quantum dynamics: \\ method and numerical tools} 


Assuming that the exciton is initially in the state $|L_B)$ uniformly delocalized over the peripheral sites, its transport across the graph is thus described by the Hamiltonian  $\mathcal{\hat{H}}$ (Eq.~\ref{eq:reduced_hamiltonian}). To simulate the associated dynamics, {Python} codes have been developed to numerically encode and diagonalize $\mathcal{\hat{H}}$ that can then be  written as 
\begin{equation}
    \mathcal{\hat{H}}  = \sum_\mu \bar{\omega}_\mu  \frac{ |\varphi_\mu )( \tilde{\varphi}_\mu| }{ (\tilde{\varphi}_\mu| \varphi_\mu )  }.
    \label{eq:reduced_hamiltonian1}
\end{equation}
In Eq.~(\ref{eq:reduced_hamiltonian1}), $|\varphi_\mu)$/$|\tilde{\varphi}_\mu ) $ represents a couple of right/left bi-orthogonal eigenstates of $\mathcal{\hat{H}}$ sharing a same complex eigenvalue  
\begin{equation}
    \bar{\omega}_\mu = \omega_\mu - i \frac{\gamma_\mu}{2},
\end{equation}
with $  \omega_\mu$ and $  \gamma_\mu$ representing respectively the real energy and the decay rate of the right/left eigenstates pair $|{\varphi_\mu})$/$|\tilde{\varphi}_\mu )$~\cite{brody14}. Knowing the  eigenstates and the eigenenergies of $\mathcal{\hat{H}}$ allows us to numerically build the time evolution operator $\hat{U}(t) = \exp(-i \mathcal{\hat{H}} t)$ written as
\begin{equation}\label{eq:evolution_op}
      \hat{U}(t) = \sum_\mu \exp(-i \bar{\omega}_\mu t) \frac{ |\varphi_\mu )( \tilde{\varphi}_\mu| }{ ( \tilde{\varphi}_\mu| \varphi_\mu )}
\end{equation}

From the knowledge of both the evolution operator and the
eigenstates, different observables can be computed. First, we
focus on the absorption probability $P_A(t)$ expressed as
\begin{equation}
   P_A(t) = 1 -  \sum_{s=0}^{L_B} |  (s| \hat{U}(t) |L_B) |^2.
   \label{eq:abso}
\end{equation}
$P_A(t)$ measures the probability for the exciton to be absorbed by the trap at time $t$. Then, to characterize the absorption process at the core of the network, we introduce a characteristic absorption time $\tau$ defined as 
\begin{equation}\label{eq:abs_time}
    \tau \longrightarrow P_A(\tau) = 50\%.
\end{equation}
The absorption time represents the moment when half of the total excitonic population is absorbed by the central core. 
In practice, the determination of the absorption time $\tau$ is realized via a numerical minimization over $t$ such that
\begin{equation}
   \tau = \min_t    (  0.5 -  P_A(t) )^2,
\end{equation}
with the \textit{Nelder-Mead} method from the {\textit{scipy}} optimization package. 

%

 \section{Numerical results: characterization of the absorption process}\label{sec:numerical}
 
 In this section, we present the results of our numerical study focusing on the absorption process at the core of the extended star graph. Every simulation presented here is realized considering the hopping constant $J$ as the reference energy unit (\textit{i.e.} $J = 1$). The absorption rate on the core is fixed to a small value ($\Gamma=0.1J$). The latter is considered as a constant for all our simulations. Note that we consider only positive values for the defect energy shift $\Delta$.

 \subsection{Absorption time : evidence of a local minimum by impurity tuning  }

\begin{figure}
     \centering
     \includegraphics[width=0.95\columnwidth]{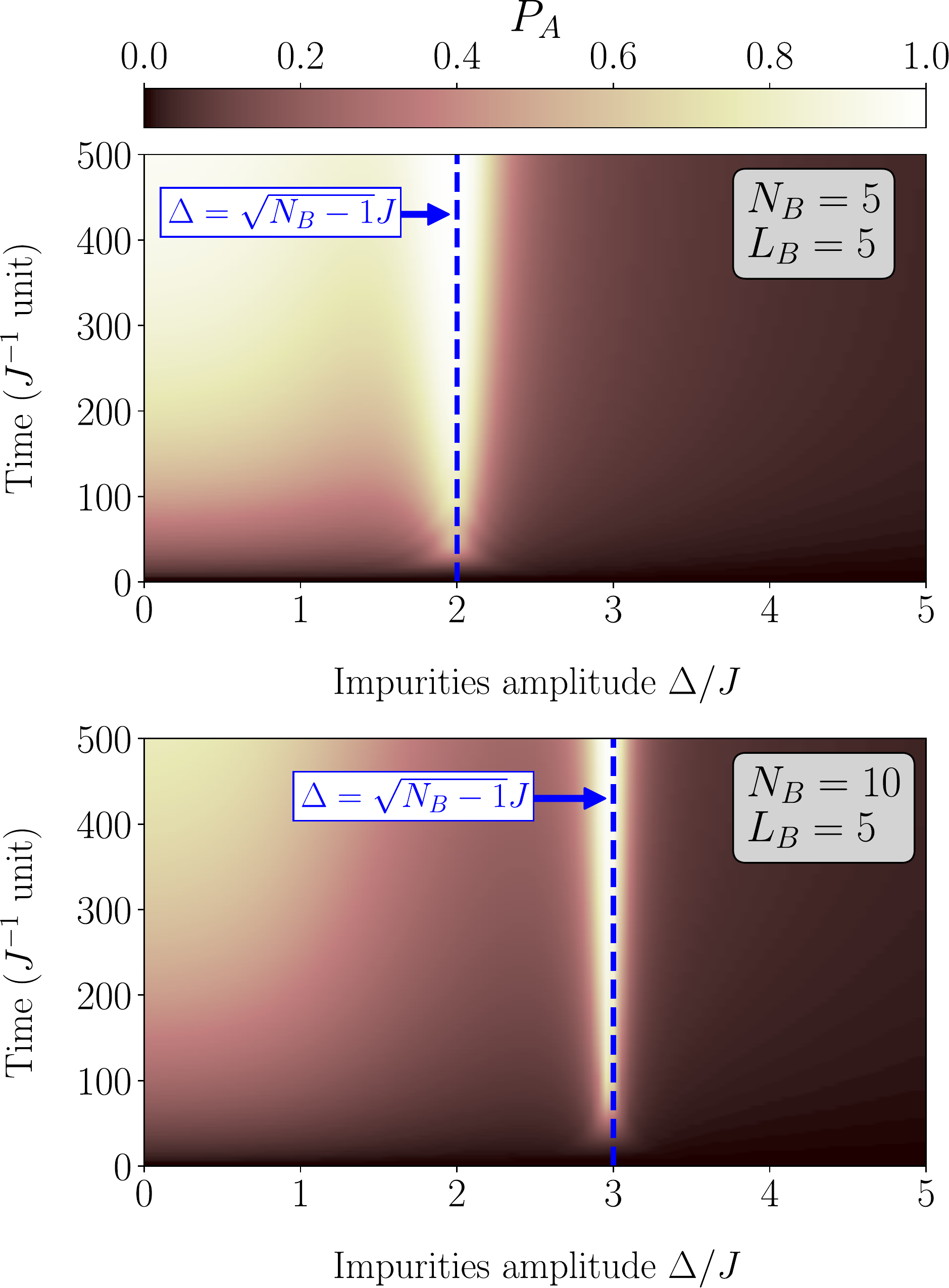} 
     \caption{\textbf{Time evolution of the absorption probability $P_A(t)$ as a function of $\Delta$ (with fixed $\Gamma=0.1J$).} \linebreak \textbf{\textit{Top panel:}} Evolution for fixed values $N_B=L_B=5$. \linebreak \textbf{\textit{Bottom panel:}} similar plot for $N_B=10$ and $L_B=5$. The vertical blue dashed lines indicate the particular value of $\Delta = \sqrt{N_B-1} J$ where a strong acceleration of the absorption process is detected. }
     \label{fig:time_evolution}
 \end{figure}

Let us focus first on the dynamics of the absorption process at the core of the star graph. To proceed, Fig.~\ref{fig:time_evolution} shows the evolution of the absorption amplitude $P_A(t)$ as a function of both the time and $\Delta$. Two sets of model parameters are considered here with $N_B=L_B=5$ (top panel) and $N_B=10$, $L_B=5$ (bottom panel).
In this figure, dark regions are associated to small absorption (\textit{i.e.} $P_A(t) \sim 0$) whereas bright regions reveal an almost complete absorption (\textit{i.e.} $P_A(t) \rightarrow 1$).
Reading the figure along the time axis (from bottom to top), we see that the absorbed population $P_A(t)$ progressively increases with the time whatever the value of $\Delta$.
However, we see that the speed of the absorption process to reach the full absorption regimes (\textit{i.e.} $P_A(t) \rightarrow 1$)   is strongly modulated by the value of $\Delta$ (as shown along the x-axis).
More particularly, when $\Delta \approx 2J$ for $(N_B=5, L_B=5)$ and when $\Delta \approx 3J$ for $(N_B=10$, $L_B=5)$, the absorption process appears to be suddenly accelerated: $P_A(t)$ reaches very quickly the full absorption limit as shown by the arising of a vertical bright beam in Fig.~\ref{fig:time_evolution}. In fact, after carrying out several simulations, it turns out that this speedup effect arises when $\Delta$ reaches a critical value $\Delta^*$ approximately equal to  $\Delta^* \approx \sqrt{N_B-1}J$. We illustrated in Fig.~\ref{fig:time_evolution} this particular value with vertical blue dashed lines.


\begin{figure} 
     \centering
     \includegraphics[width=0.9\columnwidth]{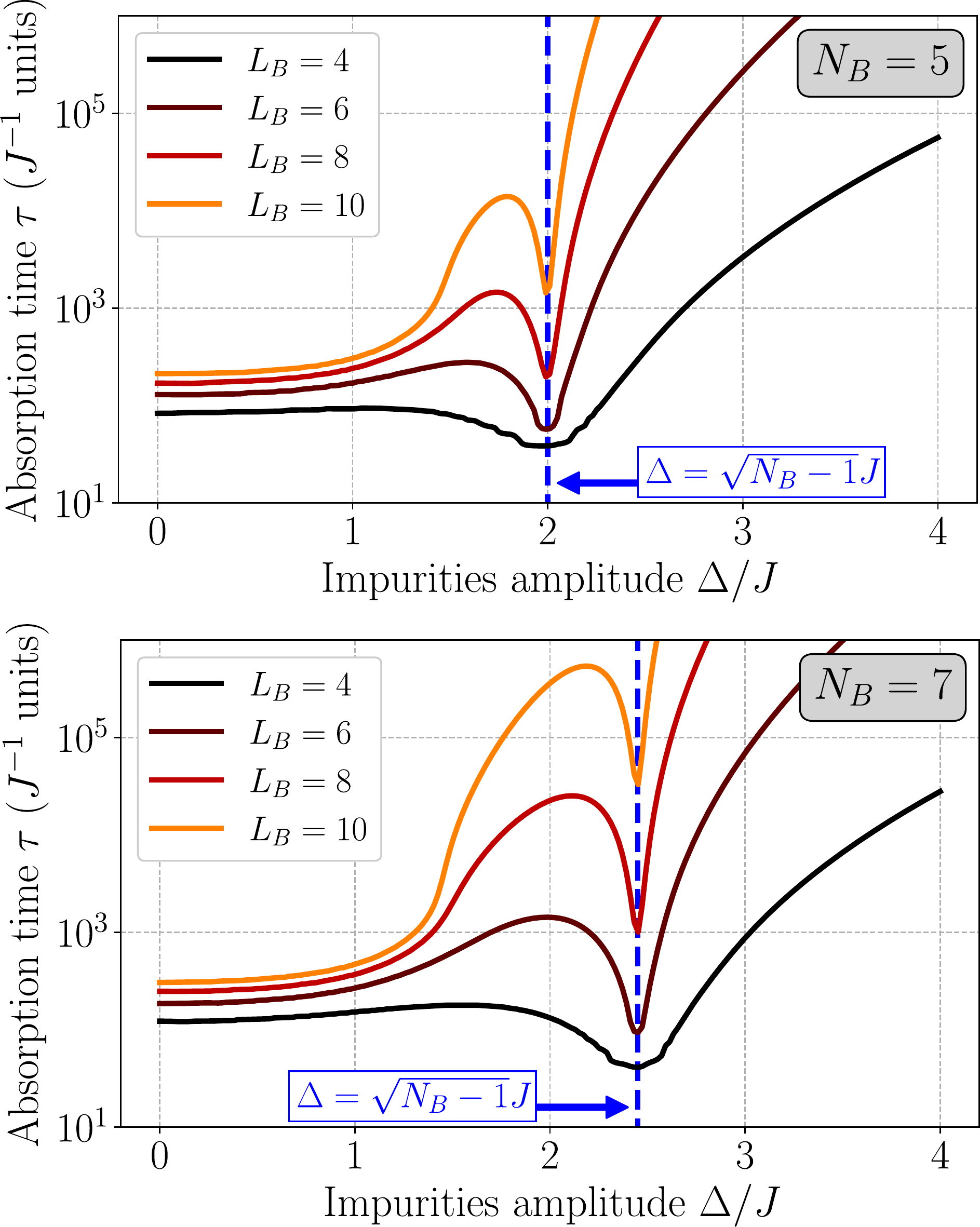}
     \caption{\textbf{Evolution of the absorption time $\tau$ with  $\Delta$, $N_B$ and $L_B$ ($\Gamma=0.1$).}  \textbf{\textit{Top panel:}} Evolution of $\tau$ as a function of $\Delta$ for a fixed value of $N_B=5$ branches and increasing values of branch length $L_B=4,6,8$ and $10$. \textbf{\textit{Bottom panel:}} similar plot for a different fixed number of  branches $N_B=7$. The vertical blue dashed lines indicate the particular value of $\Delta = \sqrt{N_B-1}J$ where a strong acceleration of the absorption process is detected.  }
     \label{fig:absorption_time}
 \end{figure}
 
To further characterize the particular behavior occurring around $\Delta^*$, let us now focus on the characteristic absorption time $\tau$ defined by Eq.~(\ref{eq:abs_time}).
Its $\Delta$ dependence is shown in Fig.~\ref{fig:absorption_time} for two different branch numbers $N_B=5$ (left panel) and $N_B=7$ (right panel) and for increasing values of the branch length $L_B=4$, 6, 8 and $10$.
Here, we see that the absorption time $\tau$ presents a  three-phase evolution. In a first phase, when $\Delta$ increases in the domain $ [0, \Delta^*[$, a progressive increasing of $\tau$ is observed.
  The slope of this increase depends on the value of $L_B$: the larger $L_B$ is, the larger is the increase of $\tau$ with $\Delta$. 
 Then, in a second phase, an abrupt change occurs when $\Delta \rightarrow \Delta^*$.
  Around this point (indicated with vertical blue dashed lines), the absorption time $\tau$ suddenly decreases to reach a local minimum.
  This feature reveals the important role of the number of branchs $N_B$ in the arising of this sudden behaviour. 
  After reaching this local minimum, the increasing of $\Delta $ in the last domain $ ] \Delta^*,+\infty[$ leads to the increasing of the absorption time $\tau $ to very large values demonstrating there that the absorption process is strongly slowed.
   As a final remark on Fig.~\ref{fig:absorption_time}, note in both panels  that all the curves are on top of each other and never cross.
   This shows that, for any fixed values of $N_B$ and $\Delta$,  increasing $L_B$ always generates a larger absorption time $\tau$. 
    This is a feature that we observed in every simulation conducted.

     \begin{figure}
     \centering
     \includegraphics[width=\columnwidth]{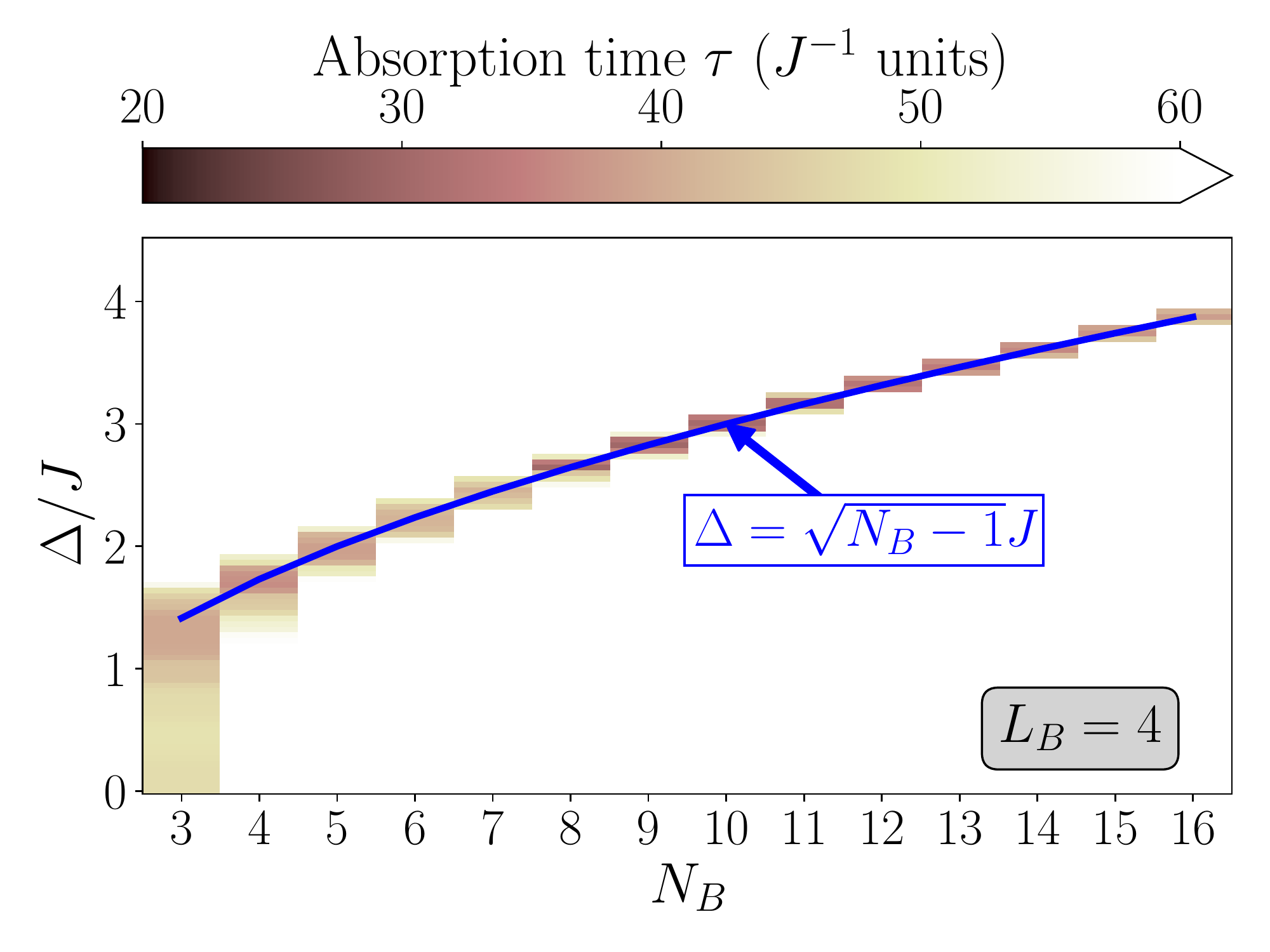}
     \caption{\textbf{Evolution of the absorption time $\tau$ with  $\Delta$ and $N_B$  ($\Gamma=0.1$).} Heatmap of the characteristic absorption time $\tau$ as a function of $\Delta$ and $N_B$ for a fixed value $L_B=4$. The  blue curve illustrate the function $\Delta = \sqrt{N_B-1}J$ which delimits a valley of local minima for the absorption time $\tau$. }
     \label{fig:systematic_absorption_time}
 \end{figure}
To finalize our analysis of the absorption time, let us show that the local minimum of $\tau$ occurring around $\Delta^*$ is actually a systematic feature of the system under study. To proceed, we illustrate in Fig.~\ref{fig:systematic_absorption_time} this feature  with a heatmap of the evolution of the characteristic absorption time $\tau$ as a function of both parameters $\Delta$ and $N_B$ (for a fixed value $L_B=4$). In this plot, we represent the function $\Delta/J = \sqrt{N_B-1}$ by the full blue curve. As shown here, the blue curve defines with a high level of precision the evolution of the valley of local minima of the absorption time $\tau$ across the parameter space ($N_B,L_B$). Note that similar trends were always observed whatever the value of $L_B$ we considered in our simulation (not shown here). 

 

 \subsection{Assessing absorption speedup in the $(N_B,L_B)$-parameter space }

 The systematic presence of the local minimum for $\tau$ suggests a potential enhancement of the absorption process when $\Delta = \Delta^*$. 
 Based on this, it is interesting to observe that, in some cases, this local minimum can go even below the absorption time obtained in the absence of defects in the periphery of the star (\textit{i.e.} $\Delta=0$). Numerical evidences of this feature are given in both panels of  Fig.~\ref{fig:absorption_time}. In these plots, we see indeed that for $L_B=4$ (black curves) the absorption time  $\tau$  actually reaches a (not only local) global minimum for $\Delta = \Delta^*$. However, this is not the case anymore when considering for example a parameter $L_B=10$ (orange curves in both panels of Fig.~\ref{fig:absorption_time}). 
 
 These observations suggest then the existence of a potential ``\textit{speedup}'' for the absorption process which can strongly depend on the architecture of the network.
As a consequence,  one can legitimately wonder: for which type of network could we expect the occurrence of an absorption speedup when $\Delta \rightarrow \Delta^*$?


To assess this question, we introduced a measure $\mathcal{S}$ of the speedup defined as 
 \begin{equation}\label{eq:speedup}
    \mathcal{S} = 1 - \frac{ \tau( \Delta^*  ) }{ \tau( \Delta=0 ) }.
\end{equation}
The measure $\mathcal{S}$ allows to estimate the reduction (or augmentation) of the absorption time obtained in the presence of defects tuned like $\Delta = \Delta^*$ compared to the case when no defects is considered $\Delta=0$. By definition,  this measure  $\mathcal{S}$ evolves on the domain $\mathcal{S} \in ]-\infty,1]$. The closer $\mathcal{S}$ gets to $1$, the more important is the speedup. Conversely, $\mathcal{S}<0$ indicates that no speedup is produced at all, namely: the absorption time $\tau$ at $\Delta=\Delta^*$ is just a local minimum in the $\Delta$-space.

 \begin{figure}
     \centering 
      \includegraphics[width=\columnwidth]{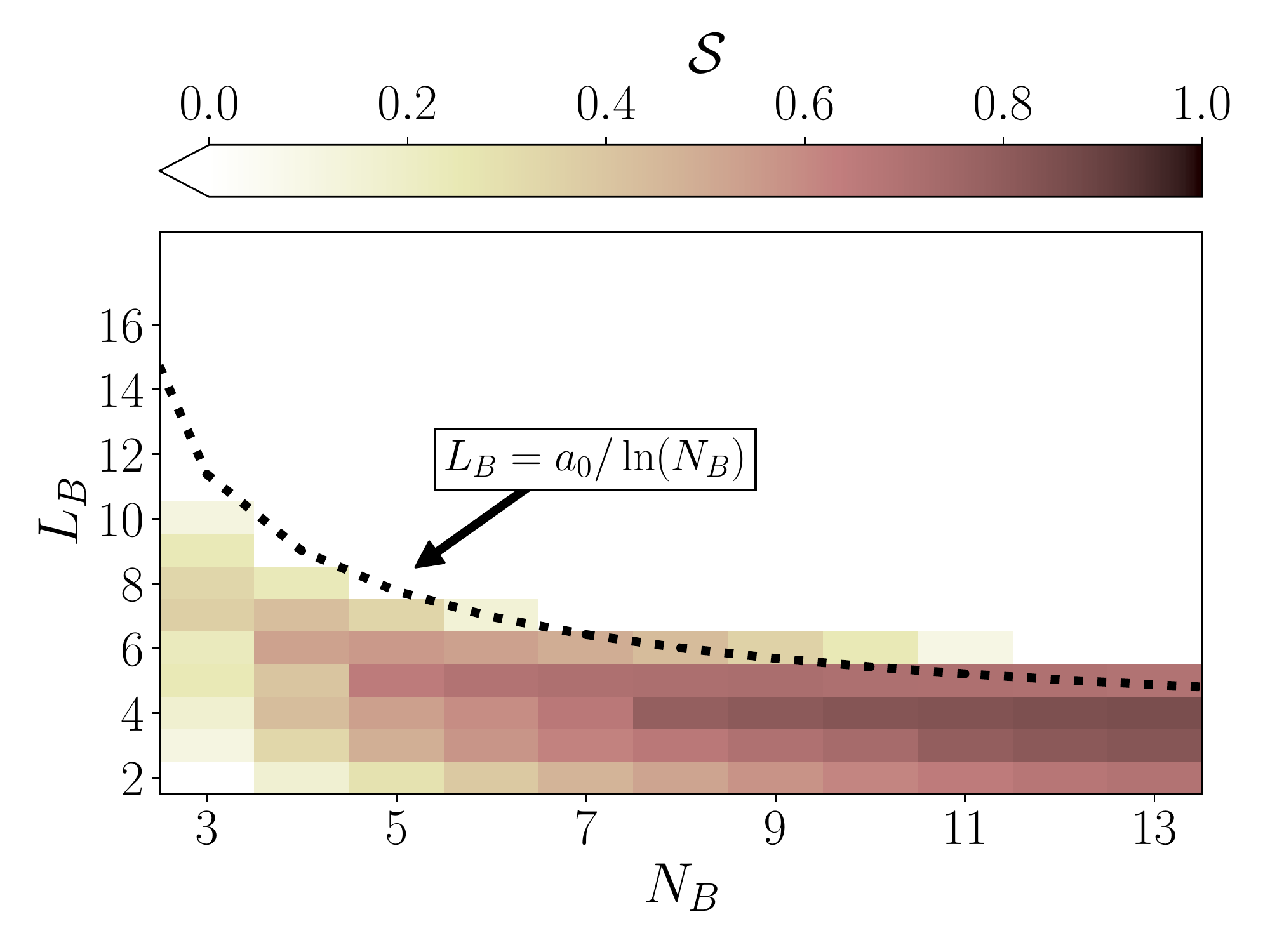} 
     \caption{\textbf{Speedup $\mathcal{S}$ of the absorption process in the $(N_B,L_B)$ parameter space.} The heatmap shows the evolution of the speedup $\mathcal{S}$ (as defined in Eq.~(\ref{eq:speedup}) in the parameter space ($N_B,L_B$). Dark colors shows where a speedup is produced (\textit{i.e.} $\mathcal{S} > 0$), whereas white area reveal where there is none (\textit{i.e.} $\mathcal{S} \leq 0$). The dotted black curve is description of the logarithmic limit defining the region where a speedup is produced. The function used is $L_B=a_0 /\ln(N_B)$ with the numerical coefficient $a_0=12.5$.}
     \label{fig:ratio}
 \end{figure}

 Fig.~\ref{fig:ratio} shows a heatmap of $\mathcal{S}$ as a function of the structure parameters $L_B$ and $N_B$. 
 On this figure, we deliberately choose to rescale the colormap to only highlight regions where a reduction is produced (\textit{i.e.} where $\mathcal{S}>0$) with a gradient of dark color.
 White regions are then related to parameters subspaces where no reduction is produced for the absorption time (when $ \Delta =\Delta^* $). 
 The results presented reveal the existence of a region in the parameter space ($N_B,L_B$) where a strong speedup can be expected.
This region is $N_B$-dependant and can be defined by the following approximate condition:  
 \begin{equation}\label{eq:LB_crit}
     \mathcal{S}>0 \Longrightarrow L_B \leq L_B^*  \text{ with } L_B^* \approx \frac{a_0}{\ln(N_B)},
 \end{equation}
where $a_0 = 12.5$ is a numerical coefficient. Thus, we see here that the parameter $L_B$ is the main limiting factor concerning the arising of a speedup in the absorption process at the core of the network (with a logarithmic scaling in $N_B$).

\subsection{ Excitonic eigenstates properties } 

To better understand the limitation of the absorption speedup with $L_B$, we will focus in this final numerical section on a key ingredient of the quantum dynamics: the excitonic eigenstates. 
As indicated by Eq.~(\ref{eq:evolution_op}), the eigenstates drive the evolution of the quantum transport inside the system through the time evolution operator.
They are thus encoding a precious information on the capacity of the exciton to efficiently delocalize to the absorbing central trap of the network.

Numerical investigations have revealed that only the two highest energy excitonic eigenstates are strongly affected by $\Delta$, and this whatever the size of the system (\textit{i.e.} the values of $N_B$ and $L_B$). 
For sake of conciseness, and because they will play a central role in our future analysis (see Sec.~\ref{sec:discussion}), we will here only focus on these two states in our discussion.

\begin{figure} 
     \centering 
     \includegraphics[width=0.9\columnwidth]{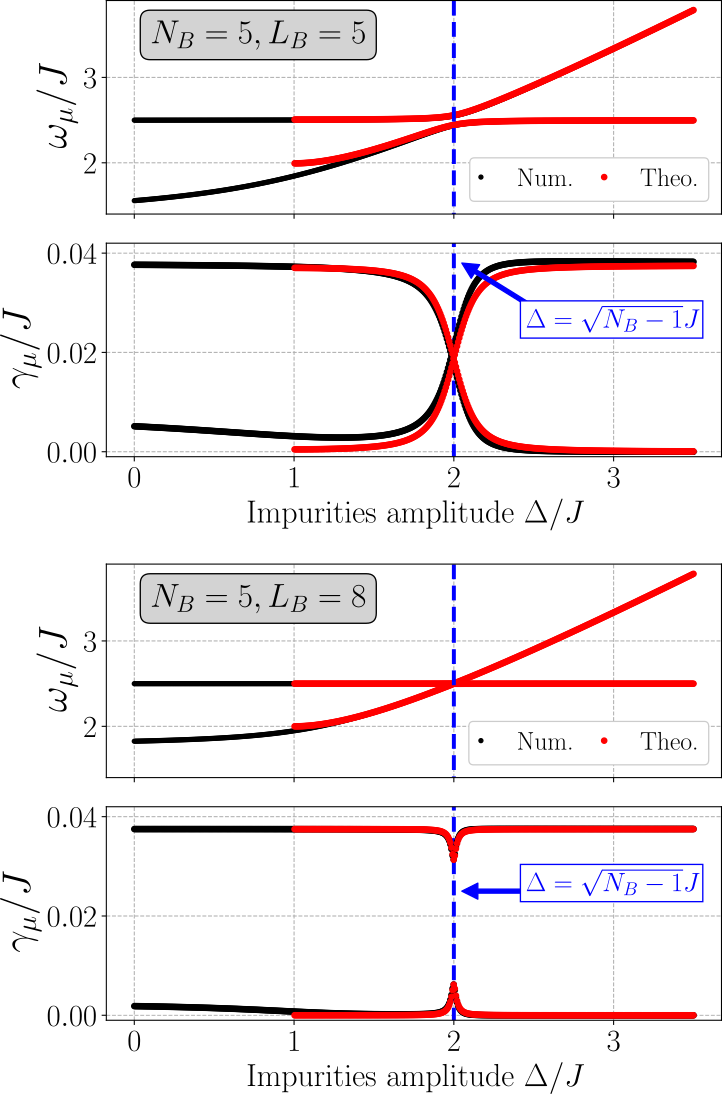} 
    \caption{\textbf{Evolution of the spectrum of the two upper-band excitonic eigenstates of $\hat{\mathcal{H}}$ as a function of $\Delta$ (with $\Gamma=0.1$).} \textbf{\textit{Top panel}}: energy $\epsilon_\mu$ and decay rate $\gamma_\mu$ of the two eigenstates for $N_B=L_B=5$. \textbf{\textit{Bottom panel}}: similar plot for $N_B=5,L_B=10$. The vertical blue dashed lines indicate the particular value of $\Delta = \sqrt{N_B-1}J$ where a strong acceleration of the absorption process is detected.}
     \label{fig:spectra} 
 \end{figure}

In Fig.~\ref{fig:spectra}, we show the evolution of the real and imaginary part of the energy of these two particular eigenstates as a function of $\Delta$.
As readily seen here, in the case $N_B=L_B=5$ (see top panel of Fig.~\ref{fig:spectra}), their real energies $\omega_\mu$  exhibit an avoided crossing around the specific value $\Delta^* \approx \sqrt{N_B-1} J$.
Simultaneously, the associated decay rates also cross in this region thus indicating that both eigenstates share a similar life time (and become quasi-degenerate).
However these features change in the case $N_B=5,L_B=8$ (see bottom panel of Fig.~\ref{fig:spectra}). 
Here, the real energies simply cross each other and not evident avoided crossing is present.
Similarly, the associated decay rates do not cross each other but simply present little bumps around $\Delta^*$. 
The two eigenstates do not share a same life time amplitude anymore: one will always be long-lived (small decay rate) whereas the second one will be short-lived (large decay rate).




Let us now focus on a structural analysis of the two upper-band eigenstates to characterize their response to variations of $\Delta$.
To proceed we will use a measure of spatial delocalization, called the Inverse Participation Ratio (IPR), which is defined like
\begin{equation}
    \text{IPR}\big( |{\varphi_\mu}) \big) = \left( \sum_{s=0}^{L_B} |(s|\varphi_\mu )|^4\right)^{-1}.
\end{equation}

Within this definition,  the IPR of an eigenstate fully localized on
one site (of the effective chain, see Figs.~\ref{fig:star}(b) and (c)) is equal to 1. By contrast, the IPR of a state uniformly
delocalized over the whole chain is equal to $L_B + 1$ (the total length).

 \begin{figure} 
     \centering 
     \includegraphics[width=0.95\columnwidth]{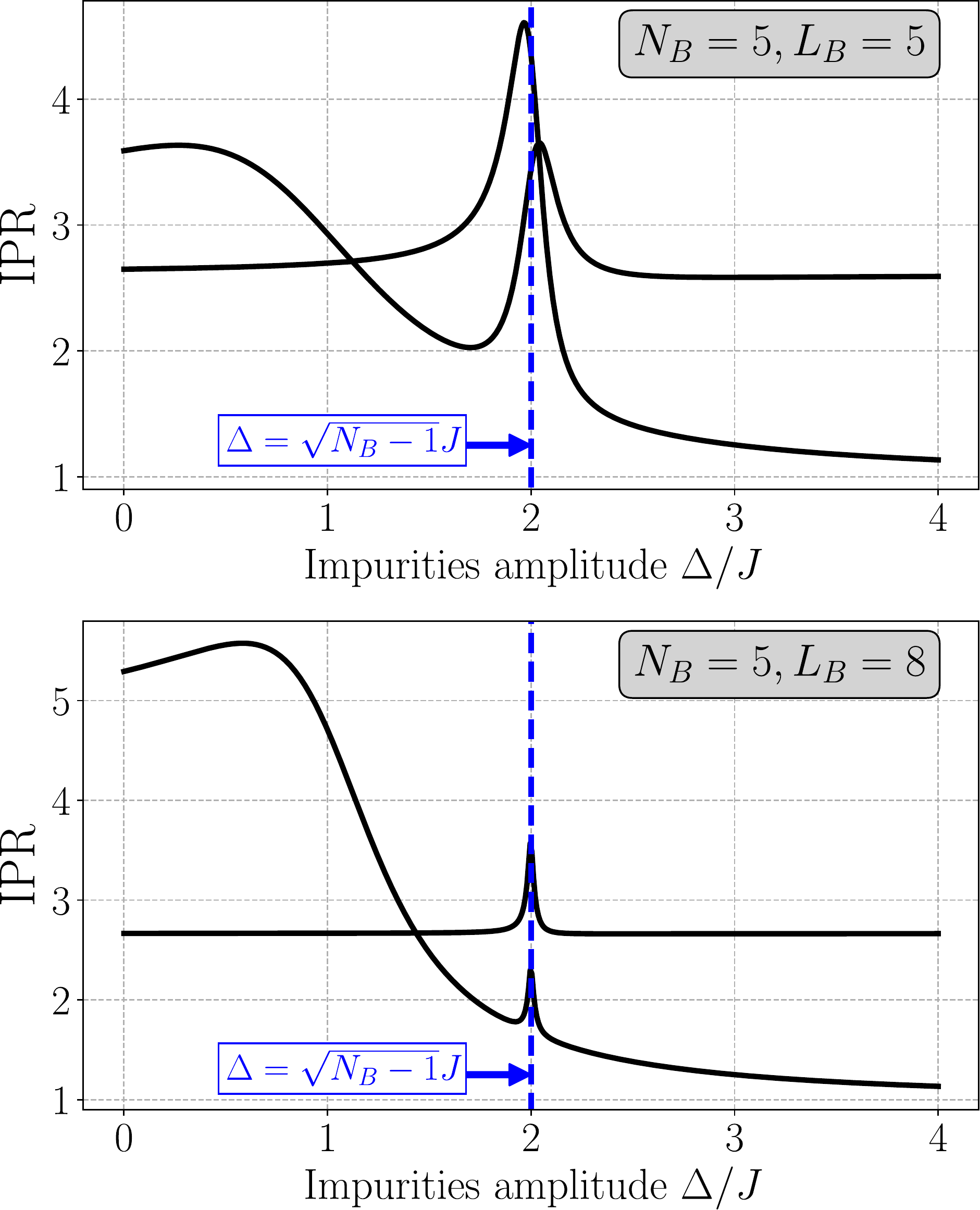} 
     \caption{ \textbf{Evolution of the Inverse Participation Ratio (IPR) of the two upper-band excitonic eigenstates of $\hat{\mathcal{H}}$ as a function of $\Delta$ (with $\Gamma=0.1$). } \textbf{\textit{Top panel}}: IPR of the two eigenstates for $N_B=L_B=5$. \textbf{\textit{Bottom panel}}: similar plot for $N_B=5,L_B=8$. The vertical blue dashed lines indicate the particular value of $\Delta = \sqrt{N_B-1}J$ where a strong acceleration of the absorption process is detected.}
     \label{fig:IPR}
 \end{figure}
 
We illustrate in Fig.~\ref{fig:IPR} the evolution of the IPR of both eigenstates as a function of $\Delta$. 
Here, two different chain lengths are considered with $L_B=5$ (top panel) and  $L_B=8$ (bottom panel) with a same  number of branches $N_B=5$. 
Focusing first on the case $L_B=5$ (left panel), we observe that a strong state restructuring occurs around $\Delta^*$ as suggested by the narrow peaks in the IPRs of both states. 
Around this particular value, the spatial spreading of the two eigenstates is maximized. 
This feature tends however to disappear when increasing the length $L_B$ of the chain as shown in the right panel of Fig.~\ref{fig:IPR}.
In this second case, the two peaks arising in the IPR of the two eigenstates are still present but strongly minimized.
Extending our numerical investigations, we observed that increasing the length of the chain tends to totally extinguish the IPR bumps in this region (not shown here). 
This feature taking place rapidly as soon as we consider $L_B > L_B^* $ (using Eq.~(\ref{eq:LB_crit}), we obtain here $L_B^* \approx 8$ for $N_B=5$ ) which follows the evolution of the absorption speedup measured in Fig.~\ref{fig:ratio}.

 
 In the coming section, we will discuss the origin of the absorption process speedup and why this phenomenon is related to the $N_B$ value. We will also explain why the structure parameter $L_B$ is a limiting factor for the occurrence of this speedup through the concept of quantum resonance from the core and the periphery of the network.
  
\section{Discussion: interpretation of the absorption process}\label{sec:discussion}

\subsection{Schrodinger equation and eigenstates}

In the previous section, numerical simulations have been carried out to characterize the excitonic absorption at the core of an extended star whose peripheral sites are occupied by defects. These defects, whose energy is shifted by an amount $\Delta$, play a key role in the efficiency of the excitonic transfer between the periphery and the core. Indeed, we have shown that when $\Delta$ reaches a specific value $\Delta^*$, a sudden speedup of the absorption process may arise, depending on the architecture of the graph. This speedup effect takes place provided that the branch length remains smaller than a $N_B$-dependant critical value approximately equal to $L_B^* \approx 12.5/\ln(N_B)$. It turns out that $\Delta^*$ only depends significantly on the branch number and it typically scales as $\Delta^* \approx \sqrt{N_B-1}J$.
Our numerical results revealed that this efficient pathway for the energy transfer originates in the restructuring of the two highest energy exciton eigenstates. When $\Delta$ is judiciously chosen, \textit{i.e.}, when $\Delta=\Delta^*$ these states delocalize to create a bridge between the periphery and the core of the star.   

To interpret and discuss these observed features, let us characterize the eigenstates of the Hamiltonian $\hat{\mathcal{H}}$ (given in Eq.~(\ref{eq:reduced_hamiltonian})). To proceed, the Schrodinger equation is expressed as 
\begin{eqnarray}
(\omega_0-i\Gamma/2) \varphi(0)+\sqrt{N_B} J \varphi(1)&=&\bar{\omega} \varphi(0) \nonumber \\
\sqrt{N_B} J \varphi(0)+\omega_0 \varphi(1)+J \varphi(2)&=&\bar{\omega} \varphi(1) \nonumber \\
... &=& ... \nonumber \\
J \varphi(s-1)+\omega_0 \varphi(s)+J \varphi(s+1)&=&\bar{\omega} \varphi(s) \nonumber \\
... &=& ...\nonumber \\
J \varphi(L_B-1)+(\omega_0+\Delta) \varphi(L_B)&=&\bar{\omega} \varphi(L_B), 
\label{eq:system}
\end{eqnarray}
where $\varphi(s)  = (s|\varphi) $ is the exciton wave function with the specific eigenenergy $\bar{\omega}$.

\subsubsection{Extended states}
 
According to the standard properties of the tight-binding model~\cite{desjonqueres96,li13}, it is straightforward to show that the finite chain shown in Fig.~\ref{fig:star}(c) supports extended states. 
Indeed, far from the side sites, the Schrodinger equation reduces to that of a linear chain with translational invariance.
Therefore, when one considers the presence of the boundaries, the Hamiltonian exhibits eigenstates that correspond to superpositions of forward and backward traveling waves as
\begin{equation}
\varphi(s) = \beta^{(+)} e^{iqs}+\beta^{(-)} e^{-iqs},
\end{equation}
where the amplitude $\beta^{(\pm)}$ can be found by applying boundary conditions. These waves, with a real wave vector $q$, define traveling states whose eigenenergies $\bar{\omega}_q=\omega_0+2J \cos(q)$ belong to the energy band $[\omega_0-2J,\omega_0+2J]$. They describe excitonic states uniformly delocalized over the branches of the star and that propagate along that branches.

\subsubsection{Localized states}

Since the finite size chain Fig.~\ref{fig:star}(c) exhibits defects which break the symmetry of the problem, the Hamiltonian $\hat{\mathcal{H}}$ supports additional states whose properties strongly differ from those of the traveling waves. These states correspond to wave functions that localize in the neighborhood of the defects and whose energies lie outside the continuous band~\cite{desjonqueres96,li13}. 

To illustrate this feature, let us seek the general solution of the Schrodinger equation as
\begin{equation}
\varphi (s) = \left\{
      \begin{aligned}
        \alpha  &     \ \ \textrm{if} & s=0 \\
        \beta^{(+)} e^{iqs}+\beta^{(-)} e^{-iqs}  & \ \ \textrm{if} & s\geq 1. \\
      \end{aligned}
    \right.
\label{eq:solloc}
\end{equation}
By inserting this solution in the Schrodinger equation far from the side sites, it turns out that the eigenenergy satisfies the dispersion relation of the infinite chain $\bar{\omega}=\omega_0+2J\cos(q)$. However, the value of the wave vector $q$ is still unknown at this stage. To determine the allowed wave vector, one must study the Schrodinger equation for $s=0$, $s=1$ and $s=L_B$. 
%
%
%
%
One obtains a system of three equations for $\alpha$, $\beta^{(+)}$ and $\beta^{(-)}$. This system exhibits non-trivial solutions if its determinant vanishes. This condition gives rise to the mode equation~\cite{pouthier97}, \textit{i.e.}, the equation whose solutions specify the allowed values of the wave vectors. It is defined as 

\begin{equation}\label{eq:mode_equation}
\frac{ \left(\dfrac{\Delta}{J}-e^{-iq}\right)\left(N_B-1-i\dfrac{\Gamma}{2J} e^{-iq}-e^{-2iq}\right)}
{\left(\dfrac{\Delta}{J}-e^{iq}\right)\left(N_B-1-i\dfrac{\Gamma}{2J}e^{iq}-e^{2iq}\right)}=e^{2iqL_B}.
\end{equation}

In a finite size chain, the mode equation cannot be solved analytically. Nevertheless, as shown in the following, it can be used to introduce relevant approximations and consequently to understand the numerical observations

To proceed, the main idea consists in a two-step approach in which one first treats independently the localization on each side of the chain by considering the limit $L_B \rightarrow \infty$. In doing so, it will be shown that each extremity of the chain exhibits localized states characterized by a complex wave vector $q=Q+i\kappa$. In that case, the right hand side of the mode equation Eq.~(\ref{eq:mode_equation}) vanishes so that the allowed $q$ values correspond to the solutions of two distinct equations written as
\begin{eqnarray}
(N_B-1)-i\frac{{\Gamma}}{2J} e^{-iq}-e^{-2iq}&=&0 \label{eq:mode0} \\
\frac{\Delta}{J}-e^{-iq}&=&0. \label{eq:modeLB}
\end{eqnarray}
Eq.~(\ref{eq:mode0}) describes states localized in the neighborhood of the sites $s=0$ and $s=1$, whereas Eq.~(\ref{eq:modeLB}) characterizes localized states occurring near the site $s=L_B$. 

The second step of our approach consists in considering finite $L_B$ values for which a hybridization may occur between states localized over different sides.

\subsection{Localization near the core of the star}

\begin{figure}[!h]
\centering
\includegraphics[width=4.5cm]{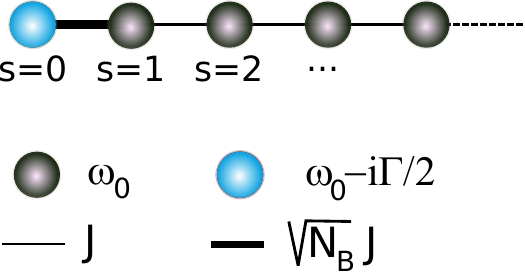}
\caption{\textbf{Illustration of the right-side semi-infinite chain model.} Here, a trap is located on site  $s=0$  and a hopping defect is connecting sites $s=0$ and $1$.}
\label{fig:loccore}
\end{figure}

To understand the occurrence of localized states in the neighborhood of the extremity sites $s=0$ and $s=1$, let us consider the semi-infinite chain displayed in Fig.~\ref{fig:loccore}. 

By solving Eq.~(\ref{eq:mode0}), provided that $N_B$ is larger than $2$, it is straightforward to show that the semi-infinite chain exhibits two localized states. The first state called ``$\varphi_{+}^{(\Gamma)}$'' is located above the continuous band whereas the second state called ``$\varphi_{-}^{(\Gamma)}$'' is located below the continuous band. The corresponding eigenenergies are defined as
\begin{equation}
\bar{\omega}^{(\Gamma)}_\pm=\omega_0\pm  \frac{J N_B}{\sqrt{N_B-1}}\sqrt{1-\frac{\epsilon^2}{N_B-1}}-i\epsilon J \frac{N_B-2}{N_B-1},
\end{equation}
where $\epsilon=\Gamma /4J$. It turns out that the unstable nature of the core site contaminates the energies of the localized states $\bar{\omega}^{(\Gamma)}_\pm=\omega^{(\Gamma)}_\pm-i\gamma^{(\Gamma)}/2$ that become complex. 
Regarding the energies $\omega^{(\Gamma)}_{\pm}$, the presence of the trap is responsible for a shift. These energies are pushed back towards the continuous band edge as $\Gamma$ increases. Nevertheless, they lie outside the band provided that $N_B>2$. They are defined as
\begin{equation}\label{eq:omgamma}
\omega^{(\Gamma)}_\pm=\omega_0\pm  \frac{ J N_B}{\sqrt{N_B-1}}\sqrt{1-\frac{\Gamma^2}{16 J^2 (N_B-1)}}.
\end{equation}
The two states ``$\varphi_\pm^{(\Gamma)}$'' are characterized by the same decay rate $\gamma^{(\Gamma)}$ defined as
\begin{equation} \label{eq:petitgamma}
\gamma^{(\Gamma)}=\frac{\Gamma}{2} \left( \frac{N_B-2}{N_B-1} \right).
\end{equation}
This decay rate, that remains smaller than that of the core site, depends on the branch number. It increases as $N_B$ increases and it converges towards the limiting value $\gamma^{(\Gamma)}_\infty=\Gamma/2$ as $N_B$ tends to infinity. Note that $\gamma^{(\Gamma)}=0$ when $N_B=2$. 
Finally, the states ``$\varphi_\pm^{(\Gamma)}$'', whose corresponding wave functions localize on the extremity sites $s=0$ and $s=1$, are characterized by the same localization length $\xi_\pm=1/\kappa$ expressed as
\begin{equation}\label{eq:xi1}
\xi_\pm=\frac{2}{\ln(N_B-1)}.
\end{equation}

\subsection{Localization near the periphery of the star} 

\begin{figure}[!h]
\centering
\includegraphics[width=6cm]{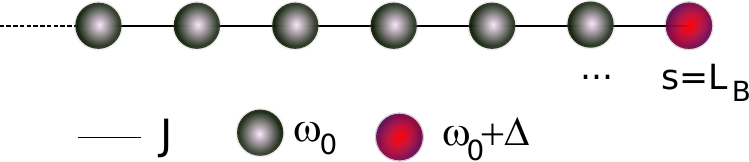}
\caption{\textbf{Illustration of the left-side semi-infinite chain model.} Here, we consider the presence of an energetic defect on the right extremity site $s=L_B$.}
\label{locper}
\end{figure}

To understand the occurrence of localized states in the neighborhood of the other extremity site $s=L_B$, let us consider the semi-infinite chain displayed in Fig.~\ref{locper}. It exhibits an energetic defect on the side site.

According to Eq.~(\ref{eq:modeLB}), provided that $\Delta$ is larger than $J$, the semi-infinite chain exhibits one localized state called ``$\varphi^{(\Delta)}$'' and whose energy is located above the continuous band ($\Delta>0)$. This energy is defined as
\begin{equation} \label{eq:omdelta}
\bar{\omega}^{(\Delta)}=\omega^{(\Delta)}=\omega_0+\frac{\Delta^2 + J^2}{\Delta}.
\end{equation}
This state whose wave function localizes near the side site $s=L_B$, is characterized by the localization length $\xi^{(\Delta)}$ expressed as
\begin{equation}\label{eq:xi2}
\xi^{(\Delta)}=\frac{1}{\ln(\Delta/J)}.
\end{equation}

\subsection{Hybridization process}


Back to the general case by considering finite $L_B$ values, let us imagine the following situation. Provided that $N_B>2$, the finite size chain will support at least two localized states whose wave functions, slightly perturbed by the boundary $s=L_B$, are localized on the left extremity of the chain. If $\Delta>J$, one expects the occurrence of a third localized state whose wave function is important in the neighborhood of the right extremity of the chain. Therefore, if the energy of that state becomes resonant with that of a state localized on the left side, a hybridization will arise. This will give rise to the occurrence of a superposition of states localized on both sides of the chain. Depending on the parameter values, this hybridization could favor an efficient energy transfer between the periphery and the core of the graph, as observed numerically in Sec.~\ref{sec:numerical}.

To illustrate this feature, let us consider that $\Delta$ is chosen so that the state ``$\varphi^{(\Delta)}"$ localized near $s=L_B$ is almost resonant with the state ``$\varphi_+^{(\Gamma)}$'' localized near the trap, \textit{i.e.} $\omega^{(\Delta)} \approx \omega^{(\Gamma)}_+$. For a finite $L_B$ value, both states interact through the overlap of their wave functions~\cite{pouthier97}. Let $C$ denotes the corresponding coupling, whose strength depends on the branch length $L_B$. In that context, if the two localized states are sufficiently far from the continuous band, the dynamics of the two highest eigenenergies is isomorphic to that of a two-level system whose Hamiltonian is defined as
\begin{equation}
\hat{H}^\text{two-level}=
\begin{pmatrix}
\bar{\omega}^{(\Gamma)}_+ & C\\
C & \bar{\omega}^{(\Delta)} 
\end{pmatrix} .
\end{equation}

The corresponding eigenvalues become complex indicating that the unstable nature of the state ``$\varphi_+^{(\Gamma)}$'' contaminates the second localized state ``$\varphi^{(\Delta)}$''. They are expressed as 
\begin{eqnarray}\label{eq:omtwolevel}
\bar{\omega}_{\pm}&=&\frac{\omega^{(\Gamma)}_++\omega^{(\Delta)}}{2}-i\frac{\gamma^{(\Gamma)}}{4} \nonumber \\
&\pm& \sqrt{\left( \frac{\omega^{(\Gamma)}_+-\omega^{(\Delta)}}{2}-i\frac{\gamma^{(\Gamma)}}{4}  \right)^2+C^2 }.
\end{eqnarray}
When one analyzes the $\Delta$ dependence of these two highest eigenenergies, two situations arise depending on the strength of the coupling $C$. 

When $L_B$ is sufficiently short when compared with the localization length of the states ``$\varphi_+^{(\Gamma)}$'' and ``$\varphi^{(\Delta)}"$, a strong coupling $C$ arises. One thus expects the occurrence of a strong hybridization process at the resonance. Nevertheless, the nature of the states of the Hamiltonian $\hat{H}^\text{two-level}$ depends on the value of $\Delta$ that controls that resonance.

For small $\Delta$ values, we are far from the resonance so that the two localized states ``$\varphi_+^{(\Gamma)}"$ and ``$\varphi^{(\Delta)}"$ remain almost independent. As observed in the top panel of Fig.~\ref{fig:spectra}, $\hat{H}^\text{two-level}$ exhibits a high energy eigenstate whose energy, approximately equal to $\omega_+^{(\Gamma)}$, is $\Delta$ independent. This state, which basically corresponds to ``$\varphi_+^{(\Gamma)}"$, is localized near the side site $s=0$ and is very sensitive to the presence of the trap. It is a short lived state characterized by an important decay rate approximately equal to $\gamma^{(\Gamma)}$. By contrast, the low energy eigenstate of $\hat{H}^\text{two-level}$ basically corresponds to ``$\varphi^{(\Delta)}"$. It is localized near the extremity site $s=L_B$ and it defines a long lived state characterized by a very small decay rate and whose energy is approximately equal to $\omega^{(\Delta)}$. In that case, the initial creation of an excitonic wave function uniformly delocalized over the periphery excites preferentially this latter long lived state. Since this state is basically localized on the defect site s$=L_B$, a rather inefficient transfer arises between the periphery and the core of the graph.  

As $\Delta$ increases, the two eigenstates of $\hat{H}^\text{two-level}$ get closer to each other. Therefore, a resonance occurs giving rise to the well-known avoided crossing effect observed in the top panel of Fig.~\ref{fig:spectra}. The resonance takes place when $\Delta$ reaches a critical value $\Delta^*$ which is the solution of the equation $\omega_+^{(\Gamma)}=\omega^{(\Delta)}$. According to both Eqs.~(\ref{eq:omgamma}) and (\ref{eq:omdelta}) it is defined as 
%
%
\begin{equation}
\Delta^* =\frac{J \left( N_B \sqrt{1-x}
+(N_B-2)\sqrt{1-\dfrac{xN_B^2}{(N_B-2)^2}} \right) }{2 \sqrt{(N_B-1)}} ,
\end{equation}
where $x=\Gamma^2/16(N_B-1)$. Note that provided that the decay rate of the trap remains small (\textit{i.e.} $\Gamma \ll  J$), this equation reveals that $\Delta^*$
mainly depends on the architecture of the star. It is  approximately equal to $\Delta^* \approx \sqrt{N_B-1} J$ in the limit $x \rightarrow 0$, in a perfect agreement with our numerical observations. 
At the resonance, the eigenvalues of the Hamiltonian $\hat{H}^\text{two-level}$ are defined as 
\begin{equation}
\bar{\omega}_{\pm}=\omega^{(\Delta^*)}\pm \sqrt{C^2- \left( \frac{\gamma^{(\Gamma)}}{4} \right)^2}-i\frac{\gamma^{(\Gamma)}}{4} .
\end{equation}
Consequently, the system supports two short lived eigenstates with the same decay rate $\gamma^{(\Gamma)}/2$, and which repel each other. These two eigenstates are symmetric and antisymmetric superpositions of the localized states ``$\varphi_+^{(\Gamma)}"$ and ``$\varphi^{(\Delta)}"$. Therefore, they correspond to delocalized states that realize a bridge between the periphery and the core of the star. This delocalization yields a sudden increase of the corresponding IPR, as observed in the top panel of Fig.~\ref{fig:IPR}, resulting in a fast and efficient energy transfer between the periphery and the core of the graph. 

As $\Delta$ increases again, one moves away from the resonance. 
One thus recovers a situation in which the two localized states ``$\varphi_+^{(\Gamma)}"$ and ``$\varphi^{(\Delta)}"$ remain almost independent. As observed in the top panel of Fig.~\ref{fig:spectra}, $\hat{H}^\text{two-level}$ exhibits a high energy eigenstate whose energy, approximately equal to $\omega^{(\Delta)}$, increases linearly with $\Delta$.
This state basically corresponds to ``$\varphi^{(\Delta)}"$ and it is  a long lived state characterized by a very small decay rate. By contrast, the low energy eigenstate of $\hat{H}^\text{two-level}$ basically corresponds to ``$\varphi_+^{(\Gamma)}"$ and it defines a short lived state whose decay rate is approximately equal to $\gamma^{(\Gamma)}$. Consequently, since the localized nature of the eigenstates recurs, the efficiency of the transfer between the periphery and the core of the star breaks down. 

At this step, let us mention that Eq.~(\ref{eq:omtwolevel}) yields a rather good estimate of both the energies and the decay rates of the two highest energy eigenstates of the star, as illustrated by the red curves shown in the top panel of Fig.~\ref{fig:spectra}. To proceed, the calculations have been carried out by using the theoretical expressions of  $\omega_+^{(\Gamma)}$ (Eq.~(\ref{eq:omgamma})), $\omega^{(\Delta)}$ (Eq.~(\ref{eq:omdelta})) and  $\gamma^{(\Gamma)}$ (Eq.~(\ref{eq:petitgamma})). For $L_B=5$ and $N_B=5$, the coupling $C=5.75\times10^{-2} J$ has been estimated by studying the energy difference at the resonance between the two highest energy eigenstates in the limit $\Gamma=0$. 

 
As the length of the branches $L_B$ increases, the overlap between the two localized states ``$\varphi_+^{(\Gamma)}"$ and ``$\varphi^{(\Delta)}"$ decreases resulting in a drastic cut in the coupling constant $C$. Far from the resonance, since the coupling $C$ does not play a significant role, the physics is the same as the one discussed previously.
By contrast, a fully different situation arises at the resonance, as displayed in the bottom panel of Fig.~\ref{fig:spectra}. In that case, the hybridization process is extremely small
between the two localized states ``$\varphi_+^{(\Gamma)}"$ and ``$\varphi^{(\Delta)}"$ so that the eigenvalues of the Hamiltonian $\hat{H}^\text{two-level}$ are now expressed as 
\begin{equation}
\bar{\omega}_{\pm}=\omega^{(\Delta^*)}-i \left( \frac{\gamma^{(\Gamma)}}{4}\pm \sqrt{ \left( \frac{\gamma^{(\Gamma)}}{4} \right)^2-C^2}  \right).
\end{equation}
In other words, it is as if the two states ``$\varphi_+^{(\Gamma)}"$ and ``$\varphi^{(\Delta)}"$ kept their own properties at the resonance. They share the same energy resulting in the disappearance of the avoided crossing process. The state almost identical to ``$\varphi_+^{(\Gamma)}"$ is mainly localized on the core of the graph. It corresponds to a short lived state whose decay rate scales as $\gamma_+ \approx \gamma^{(\Gamma)}-4C^2/\gamma^{(\Gamma)}$. By contrast the state almost identical to ``$\varphi^{(\Delta)}"$ is localized on the periphery. It thus remains quite insensitive to the trap and it corresponds to a long lived state whose decay rate is approximately equal to $\gamma_-\approx 4C^2/\gamma^{(\Gamma)}$.  As observed in our numerical simulation, the initial creation of an excitonic wave function uniformly delocalized over the periphery of the star graph excites this latter long lived state. As a result, a slow and inefficient transfer arises between the periphery and the core of the graph.  Note that the localized nature of these two states prevents their IPR to take a significant value, as displayed in the bottom panel of Fig.~\ref{fig:IPR}. Moreover, as illustrated by the red lines in the bottom panel of Fig.~\ref{fig:spectra}, Eq.~(\ref{eq:omtwolevel}) still provides a rather good estimate of both the energies and the decay rates of the two highest energy eigenstates of the star, the coupling being fixed now to $C=1.74\times10^{-3} J$ for $L_B=10$ (and $N_B=5$). 

To conclude, let us mention that the present approach shows that resonance-induced efficient energy transfer between the periphery and the core of the star requires a coupling $C$ between ``$\varphi_+^{(\Gamma)}"$ and ``$\varphi^{(\Delta)}"$ sufficiently strong.
This coupling is proportional to the overlap between the two states ``$\varphi_+^{(\Gamma)}"$ and ``$\varphi^{(\Delta)}"$.
At the resonance, both states are characterized by an almost identical localization length $\xi \approx 2/\ln(N_B-1)$ (see Eq.~(\ref{eq:xi1}) and Eq.~(\ref{eq:xi2}) for $\Delta \approx J\sqrt{N_B-1}$).
Therefore, one expects this overlap to scale approximately as $\exp(-L_B/\xi)$ so that the coupling strength depends on the competition between the branch length $L_B$ and the localization length~$\xi$.
In other words, a strong $C$ value (or a weak $C$ value) refers to a situation in which $L_B$ is shorter (or larger) than $\xi$.
This feature suggests that resonance-induced efficient energy transfer arises provided that $L_B$ is smaller than a critical value $L_B^*\sim\xi$, \textit{i.e.} a critical value that scales as $L_B^* \propto 1/\ln(N_B-1)$, in a quite good agreement with the results displayed in Fig.~\ref{fig:ratio} and the behavior given in Eq.~(\ref{eq:LB_crit}).

\section{Conclusion}



%
%
%
%
 
In this paper, we studied the quantum dynamics of a photo-excitation uniformly distributed at the periphery of an extended star graph composed of a central absorbing site connected to $N_B$ branches of length $L_B$.
We investigated the question of the energy absorption at the core of the network and how the latter can be improved by the inclusion of peripheral defects with a tunable energy amplitude $\Delta$.  

By mean of numerical/analytical developments, we demonstrated the possibility to generate a strong speedup for the energy absorption process (\textit{i.e.} a strong minimization of the absorption time).
To produce this speedup, the amplitude $\Delta$ of the peripheral energy defects should be tuned as $\Delta = \Delta^*$ with $\Delta^* \propto \sqrt{N_B-1}$. Moreover, the architecture of the star graph has to satisfy the condition $L_B \leq L_B^*$ with $L_B^* \propto 1/\ln(N_B)$. 
To interpret the arising of the speedup, analytical/numerical developments were conducted.
We then demonstrated that the origin of this feature takes place in the restructuring of the two highest energy exciton eigenstates. When $\Delta=\Delta^*$, these states delocalize to create a bridge between the periphery and the core of the star. 
This effect is important when $L_B \leq L_B^*$ and vanishes almost totally when  $L_B >  L_B^*$.

Therefore, our analysis of the excitonic dynamics made it possible to determine the structural rules governing the arising of an absorption enhancement for the extended star graph. 
Naturally, these results are preliminary and  motivate a series of questions that could represent interesting starting point for future works.
For example, it would be interesting to investigate if similar rules (for $\Delta$, $N_B$ and $L_B$) would hold in the case of a more realistic system including the presence of a perturbative environment for the photo-excitation. 
To proceed, dissipative and dephasing effects could be included in future simulations to mimic the effect of the excitonic optical recombination and the presence of an external phonon bath.
Such new ingredients will obviously perturb the excitonic dynamics and thus modify the conditions for the arising of the absorption speedup. 
Still along the lines of targeting more realistic systems, the presence of natural disorder of the site energies and hopping constants in the networks could also be considered as a way to model the imperfections naturally present in molecular networks. 
In this context, depending on the strength of the local disorder, the symmetry breaking generated would naturally open new paths for the excitonic dynamics leading to potential different absorption mechanisms. 
All these ideas are left for future projects and papers.

\phantomsection
\bibliography{biblio}


\end{document}